\documentclass[a4paper,11pt]{article}

\usepackage[utf8]{inputenc}
\usepackage[T1,T2A]{fontenc}
\usepackage{amsmath,amsfonts,amssymb}

 \usepackage{graphicx}
\usepackage{wrapfig}
\usepackage{subfig}
\usepackage{hyperref}
\usepackage[dvipsnames]{xcolor}

\begin{document}

\begin{titlepage}

\begin{center} {\LARGE \bf On gravitational preheating} \end{center}

\vspace{1cm}

\begin{center}
  {\bf Oleg Lebedev and Jong-Hyun Yoon}
\end{center}
  
\begin{center}
  \vspace*{0.15cm}
  \it{Department of Physics and Helsinki Institute of Physics,\\
  Gustaf H\"allstr\"omin katu 2a, FI-00014 Helsinki, Finland}\\
\end{center}
  
\vspace{2.5cm}

\begin{center} {\bf Abstract} \end{center}
\noindent   We consider dark matter production during the inflaton oscillation epoch.
 It is conceivable  that renormalizable interactions between  dark matter and inflaton may be negligible. In this case, the leading role is played by higher dimensional operators generated by gravity and thus suppressed by the Planck scale. We focus on dim--6 operators and study the corresponding particle production in perturbative and non--perturbative regimes. We find that the dark matter production rate is dominated by  non--derivative operators involving higher powers of the inflaton field. 
 Even if they appear with small Wilson coefficients, such
  operators   can readily account for the correct dark matter abundance.

\end{titlepage}

\tableofcontents

 \section{Introduction} 
 
 The nature of dark matter (DM) remains an outstanding mystery of modern physics. The null DM direct detection results motivate one to  explore the possibility that dark matter has feeble interactions,
 in which case it does not reach thermal equilibrium with the environment. Therefore, its abundance is sensitive to the production mechanism. One of such mechanisms is provided by gravity,
 which can efficiently produce particles in non--adiabatic environments.

 In the absence of any non--gravitational couplings, the expansion of the Universe is itself a source of particle production \cite{Parker:1969au},\cite{Grib:1976pw}. For example, the equation of motion (EOM) for a 
 momentum mode $\chi_k$ of a free scalar 
 in the Friedmann Universe with the metric $ds^2 = a^2(\eta) \, (d\eta^2 - d {\bf x}^2)$ reads \cite{Kuzmin:1998kk}
  \begin{equation}
  \chi_k^{\prime \prime} + \omega_k^2 \chi_k =0 \;,
  \end{equation}
 where 
 \begin{equation}
 \omega_k^2 = k^2 - {a^{\prime \prime} \over a} (1-6\xi ) + m_\chi^2 a^2 \;,
 \end{equation}
 $a$ is the scale factor, $\xi$ is the non--minimal coupling to gravity \cite{Chernikov:1968zm} and the prime denotes differentiation with respect to conformal time $\eta$. Time variation of $\omega_k$ is non--adiabatic if
 $\omega^\prime_k /\omega_k^2 \gtrsim 1$, which implies particle creation due to expansion. For low $k$ and conformal coupling $\xi=1/6$, this is equivalent to $a^\prime /a^2  = H \gtrsim m_\chi$ such 
 that particles lighter than the Hubble rate $H$ are constantly created.  The effect can be  even stronger for non--conformal $\xi$. The accumulated abundance of $\chi$ can constitute dark matter \cite{Kuzmin:1998kk,Chung:2001cb},
 depending on $m_\chi$ and its self--interaction.  
 
 Such particle production can be viewed in 
  terms of the scalar field condensate $\langle \chi^2 \rangle$. Light scalars are subject to quantum fluctuations of order $H$ \cite{Starobinsky:1994bd} so that a semi--classical field $\chi$ 
 experiences a random walk. As a result, a significant $\langle \chi^2 \rangle$ can accumulate, for example, by the end of inflation and play the role of dark matter \cite{Enqvist:2014zqa},\cite{Markkanen:2018gcw}.
 Again, this mechanism is purely gravitational.  The consequent dark matter distribution is not correlated with the inflaton fluctuations, therefore 
 this possibility is subject to strict isocurvature constraints \cite{Markkanen:2018gcw}.
 
 In this work, we focus on other aspects of particle production due to gravitational effects. Specifically, gravity is believed to generate couplings between different sectors of the theory as long as these are 
 consistent with gauge symmetries. The corresponding operators may be non--renormalizable and thus suppressed by the Planck scale. Nevertheless, they can play an important role in dark matter production.
 This was recently emphasised in \cite{Mambrini:2021zpp},\cite{Clery:2021bwz}, where the effects due to tree level graviton exchange were considered.
 
 We study dark matter production 
 during the inflaton oscillation phase, which sets in immediately after inflation and 
 creates a non--adiabatic environment \cite{Dolgov:1989us,Kofman:1994rk,Shtanov:1994ce}.
  Using the  effective field theory approach, we focus on the leading gravity--induced dim--6 operators assuming that the renormalizable couplings between the inflaton and dark matter
 vanish. If DM is feebly interacting, its eventual abundance is determined by the number of DM quanta produced at this ``preheating'' stage. To this end, we
  identify the dominant operator and study 
  whether it can be responsible for the correct 
   dark matter abundance.

 \section{The set--up}
 
 Consider the possibility that the renormalizable couplings between the inflaton $\phi$  and dark matter  $s$ are zero or negligibly small. Then, the  $\phi-s$ interaction can be described by a series of higher dimensional operators generated by gravity  and  thus suppressed by the Planck scale $M_{\rm Pl}$. Let us assume for simplicity that these operators exhibit an approximate $\phi \rightarrow - \phi$ symmetry such that the lowest operator dimension is six:\footnote{The dim--5 operator $\phi^3 s^2$ may in general be present, in which case it would dominate dark matter production. }
 \begin{eqnarray}
 \Delta {\cal L}_6 = {C_1 \over M_{\rm Pl}^2 } \; (\partial_\mu \phi)^2  s^2 +   {C_2 \over M_{\rm Pl}^2 } \;  (\phi \partial_\mu \phi) (s \partial^\mu s) +
 {C_3 \over M_{\rm Pl}^2 } \; (\partial_\mu s)^2 \phi ^2 - {C_4 \over M_{\rm Pl}^2 } \; \phi^4 s^2 
 - {C_5 \over M_{\rm Pl}^2 } \; \phi^2 s^4 \;, 
 \label{L}
 \end{eqnarray}
 where we have replaced the covariant derivatives with the partial ones. The inflaton field  with mass $m_\phi$ is assumed to have either (locally) quadratic or quartic potential, while the dark matter mass 
 $m_s$ is taken to be negligible compared to the typical scales of the problem.
 Some of the above  operators such as $ (\partial_\mu \phi)^2  s^2 $ and $(\partial_\mu s)^2 \phi ^2$, along with ${m_{\phi,s}^2 \over M_{\rm Pl}^2} \, \phi^2 s^2$, are generated by the tree level graviton 
 exchange \cite{Mambrini:2021zpp}.
 Others can be generated at loop level and non--perturbatively.  Since gravity is non--renormalizable,  their coefficients should be treated as arbitrary  input parameters.
   The above  interactions are responsible for dark matter production after inflation, in particular, during the inflaton oscillation phase. Depending on the $C_i$ coefficients,
  the production mechanism can be perturbative or non--perturbative (resonant). 
   
  To mention but one example, some of the above operators operators appear automatically in theories with non--minimal couplings of scalars to gravity \cite{Chernikov:1968zm}.
  In particular, 
$(\partial_\mu s)^2  \phi^2$, $(\partial_\mu \phi)^2  s^2$ and ${s^2 } \, V(\phi)$ are induced already at tree level by
 the metric transformation from the Jordan frame to the Einstein frame \cite{Salopek:1988qh}. The  (unsuppressed) operators $\phi^2 s^4$ and $\phi^4 s^2$
 appear  in these models 
   at 1--loop via the graviton loop. Their coefficients are proportional to the product of the non--minimal couplings and the loop factor, and thus  expected to be significant (in the absence of 
   symmetry arguments).
   In general, it is a challenging task to estimate 
   the relative size of the different operators since this can only be done reliably within 
   UV complete gravity theories.

On shell, two of the derivative operators can be eliminated via integration by parts:
\begin{eqnarray}
&& (\partial_\mu \phi)^2  s^2  \rightarrow  (\partial_\mu s)^2 \phi ^2 + m_\phi^2 \phi^2 s^2 \;, \\
&& (\phi \partial_\mu \phi) (s \partial^\mu s) \rightarrow - {1\over 2} (\partial_\mu s)^2 \phi^2 \;,
\end{eqnarray}
where we have neglected the dark matter mass and the Hubble rate,  which, during preheating  is  small compared to the particle energy. (We consider the Hubble--induced effects in 
Section \ref{resonant}).
Focussing on dark matter pair production, we can thus restrict ourselves to the operators 
\begin{equation}
{\cal O}_3 =  {1 \over M_{\rm Pl}^2 } \; (\partial_\mu s)^2 \phi ^2 ~~,~~ {\cal O}_4= {1 \over M_{\rm Pl}^2 } \; \phi^4 s^2 \;,
\end{equation}
amended with the renormalizable interaction
\begin{equation}
{\cal O}_{\rm renorm} =  {m_\phi^2 \over M_{\rm Pl}^2 } \;  \phi^2 s^2 \;.
\label{O-ren}
\end{equation}
Although this term is renormalizable, the coupling strength is highly suppressed: for typical inflaton masses it is below $10^{-10}$.
The operator $\phi^2 s^4$ produces a final state with 4 DM quanta. 
The corresponding reaction rate is similar to that of the derivative operator, since the final state phase space gives analogous energy dependence.
Therefore,
 we will not discuss this operator  separately within the perturbative regime, while its non--perturbative analysis will be presented in Sec.\,\ref{resonant}.

Clearly, operators ${\cal O}_3$ and ${\cal O}_4$ exhibit qualitatively different behaviour in regard to dark matter production. Indeed, ${\cal O}_3$ involves the particle energy
which is of the order of the (effective) inflaton mass, while in ${\cal O}_4$ this dependence is replaced by the inflaton field value.
The latter 
 is not far from the Planck scale in typical models, thus 
 \begin{equation}
  \phi \gg E_\phi
  \end{equation}
 and one
expects much more efficient DM production from ${\cal O}_4$. In what follows, we make this argument more quantitative.

\section{Perturbative dark matter production}

An oscillating classical background can lead to particle production \cite{Dolgov:1989us,Kofman:1994rk,Shtanov:1994ce}. After inflation, $\phi$ oscillates coherently in either $\phi^2$ or $\phi^4$ potential, depending on the inflationary model.
As a result, the $\phi-s$ couplings induce dark matter pair production. If the corresponding coupling is small, the process can be described perturbatively. Below, we consider the 
contributions of the 3 basic operators to this reaction.   We treat the Hubble expansion adiabatically such that the time dependence can be inserted in the inflaton oscillation amplitude at the end of the calculation. Also, we treat the produced dark matter particles as free and neglect backreaction. These approximations are justified at small inflaton--DM couplings.

\subsection{{$\phi^2 s^2$} interaction}

Consider the 4--point interaction
\begin{equation}
- \Delta {\cal L}_{\rm renorm} = {1\over 4} \lambda_{\phi s} \, \phi^2 s^2 \;,
\label{L-ren}
\end{equation}
where $\lambda_{\phi s} \sim m_\phi^2 / M_{\rm Pl}^2$.
Let us expand the inflaton field as
 \begin{equation}
\phi^2(t) = \sum_{n=-\infty}^\infty \zeta_n e^{-in \omega t} \;,
 \end{equation}
 where the coefficients $\zeta_n$ are time--independent.
 Creation of a two--particle DM state with momenta $p,q$ from the vacuum is described by the amplitude (in Peskin--Schroeder conventions  \cite{Peskin:1995ev})
   \begin{equation}
-i \int_{-\infty}^\infty dt \langle f | V(t) | i \rangle = - i \,{ \lambda_{\phi s} \over 2 } \, (2\pi)^4 \delta ({\bf{p}} + {\bf{q}})  \sum_{n=1}^\infty \zeta_n \delta(E_p +E_q -n \omega) \;,
 \end{equation}
 with $V(t)$  given by Eq.\,\ref{L-ren}.
 The corresponding invariant amplitude for the  
 $n$-th inflaton mode decay is ${\cal M}_n = -\lambda_{\phi s} \zeta_n /2$. The resulting reaction rate  for DM pair production per unit volume is 
    \begin{equation}
 \Gamma = \sum_{n=1}^\infty \Gamma_n =   \sum_{n=1}^\infty  {1\over 2} \int |{\cal M}_n|^2 d \Pi_n =
 {\lambda_{\phi s}^2 \over 64 \pi  } \sum_{n=1}^\infty  |\zeta_n|^2 \sqrt{1- \left(   {2m_s \over n \omega} \right)^2} \; \theta(n\omega -2 m_s) \;.
 \label{Gamma-ss}
  \end{equation}
 Here we have kept the  DM mass for generality. The inflaton decay rate $\Gamma_\phi$ can be computed using energy conservation, $\rho_\phi \Gamma_\phi  = \langle E \rangle  \Gamma   $, 
 where $\rho_\phi$ and $\langle E \rangle$ are the inflaton energy density and the average energy of the decay products, respectively.
    Hence,
  \begin{equation}
 \Gamma_\phi  =
 {\lambda_{\phi s}^2 \omega \over 64 \pi \rho_\phi  } \; \sum_{n=1}^\infty  n  |\zeta_n|^2 \sqrt{1- \left(   {2m_s \over n \omega} \right)^2} \; \theta(n\omega -2 m_s) \;.
  \end{equation}
 In the massless limit, one thus recovers the result of \cite{Ichikawa:2008ne}.

  \subsection{$\phi^2 (\partial_\mu s)^2$ interaction}

The calculation proceeds as above, except the final state receives an additional momentum--dependent factor in the amplitude:
$$       p\cdot q =  {1\over 2} (p+q)^2 = {1\over 2} (E_p +E_q)^2 = {1\over 2} n^2 \omega^2 \;,
$$
where the DM mass has been neglected.
So, effectively in the amplitude for the quartic interaction, one replaces $\lambda_{\phi s} \rightarrow 4 C_3/M_{\rm Pl}^2 \, p\cdot q $,
which in the $m_s \rightarrow 0$ limit leads to 
  \begin{equation}
 \Gamma =     {C_3^2 \, \omega^4 \over 16 \pi M_{\rm Pl}^4}   \,  \sum_{n=1}^\infty  n^4 \vert \zeta_n \vert^2 \;.
  \label{Gamma-ss-1}
  \end{equation}

 \subsection{$\phi^4 s^2$ interaction}

The calculation is similar to that for the $\phi^2 s^2$ case. $\phi^4$ can be expanded as
\begin{equation}
\phi^4(t) = \sum_{n=-\infty}^\infty \hat \zeta_n e^{-in \omega t} \;,
 \end{equation}
where
\begin{equation}
\hat \zeta_n = \sum_{m=-\infty}^\infty \zeta_{n-m} \zeta_m \;.
 \end{equation}
Replacing $\lambda_{\phi s} \rightarrow 4C_4/M_{\rm Pl}^2$, in the massless DM limit we get
 \begin{equation}
 \Gamma =   { C_4^2 \over 4\pi M_{\rm Pl}^4 } \, \sum_{n=1}^\infty  |\hat\zeta_n|^2   \;.
 \label{Gamma-ss-2}
  \end{equation}

\subsection{Relative efficiency}

Let us estimate the relative particle production efficiency of the different operators. If the inflaton potential is quadratic, 
$V(\phi) = {1\over 2} m_\phi^2 \phi^2$, we have 
\begin{equation}
\phi(t) = \phi_0 \cos m_\phi t \;,
\end{equation}
where $\phi_0$ is the oscillation amplitude. (The fact that $\phi_0$ decreases  slowly in time, $\phi_0 \propto 1/(m_\phi t)$, is insignificant for our purposes).
In the quartic case, $V(\phi) = {1\over 4} \lambda_\phi \phi^4$, the inflaton field is given by the Jacobi cosine,
 \begin{equation}
\phi(t)= \phi_0  \,{\rm cn} \left(\sqrt{\lambda_\phi} \phi_0 \, t, {1\over \sqrt{2}}\right)=
{\sqrt{\pi} \Gamma \left( {3\over 4}\right) \over   \Gamma \left( {5\over 4}\right)}\, \phi_0 \, \sum_{n=1}^\infty
 \left( e^{i(2n-1)\omega t} +       e^{-i(2n-1)\omega t}     \right) \, { e^{-(\pi/2) (2n-1)}   \over 1+ e^{-\pi (2n-1)}      }\;,
      \end{equation}
where
\begin{equation}
\omega = {1\over 2} \sqrt{\pi \over 6 } { \Gamma \left( {3\over 4}\right) \over  \Gamma \left( {5\over 4}\right)}\, m_\phi^{\rm eff}
\label{omega}
     \end{equation}
and 
\begin{equation}
  m_\phi^{\rm eff} = \sqrt{3 \lambda_\phi} \phi_0 \;.
     \end{equation}
For many purposes, the above sum can be approximated by the first term with $n=1$.

The relative efficiency of $ {\cal O}_3$  and ${\cal O}_4 $ is given by 
\begin{equation}
{\Gamma \left[  {\cal O}_3 \right] \over \Gamma \left[  {\cal O}_4 \right] } \sim {C_3^2 \over C_4^2}\, {\omega^4 \over \phi_0^4} \;.
\end{equation}
 Clearly, the reaction rate due to $\phi^2 (\partial_\mu s)^2$ is much suppressed
compared to that of $\phi^4 s^2$. For the quadratic inflaton potential, the suppression factor is 
\begin{equation}
m_\phi^4 / \phi_0^4 \sim 10^{-20} \;,
\end{equation}
assuming the typical values $\phi_0 \sim 1$ and $m_\phi \sim 10^{-5}$ in Planck units. In  the quartic case, $  \omega^4 / \phi_0^4 \sim \lambda_\phi^2 < 10^{-20}$ 
for typical $\lambda_\phi < 10^{-10}$. 

The contribution of the $\phi^2 s^2$--operator of the form (\ref{O-ren}) is similarly suppressed by $m_\phi^4/\phi_0^4$. 
It is also clear that, due to the phase space integral,  the rate of the ${\cal O}_5$--induced process 
$\phi \phi \rightarrow ssss$   contains an additional factor of $E_\phi^4 /\phi_0^4$ compared to the pair production rate from ${\cal O}_4$.
Hence we conclude that the $\phi^4s^2$ interaction dominates DM production, unless there is 
a large hierarchy in the Wilson coefficients, e.g.
 $C_3/C_4 \sim 10^{10}$.

The above calculation also tells us that higher dimensional operators
\begin{equation}
{\tilde C_6 \over M_{\rm Pl}^4} \phi^6 s^2  + {\tilde C_8 \over M_{\rm Pl}^6} \phi^8 s^2 + ...
\end{equation}
are important. Indeed, their contributions are only suppressed by $\phi_0^4/M_{\rm Pl}^4$, etc. relative to that of ${\cal O}_4 $. 
If the inflaton amplitude is not   far away from the Planck scale, this suppression is not very significant.

\section{Resonant dark matter production via dim--6 operators}
\label{resonant}

Perturbative calculations ignore the Bose enhancement of the amplitude due to the presence of  identical  states.  Depending on the coupling,  this enhancement can be very significant and lead
to resonant production  \cite{Kofman:1994rk,Kofman:1997yn,Greene:1997fu}. Such a  regime can be described semiclassically by analyzing the equations of motion for the DM field $s$.
 In what follows, we compare the corresponding 
 resonant particle production via 
 operators  $ {\cal O}_3$  and ${\cal O}_4 $.

To derive the EOM for $s$ in the presence of the inflaton background $\phi (t)$,  consider the action  
\begin{equation}
S= \int d^4 x \sqrt{| g |} \left(  {1\over 2} \, {\cal K}(\phi) \, g^{\mu \nu} \partial_\mu s \, \partial_\nu s        - {\cal V}        \right) \;,
\end{equation}
where the kinetic function ${\cal K}(\phi)$ depends on $\phi $ only and ${\cal V}$ is the $s-dependent$ part of the  scalar potential. Here $g = {\rm det} \, g_{\mu \nu}$ and 
the Friedmann metric is $ g_{\mu \nu} = {\rm diag} (1, -a^2,-a^2,-a^2)$. Since $\phi$ and $a$ are functions of time only, variation of the action with respect to $s$ yields
\begin{equation}
\ddot s - {1\over a^2} \,\partial_i \partial_i s + \left(   {\dot {\cal K} \over {\cal K}} +3H  \right)  \dot s + {{\cal V}^{\prime}_s \over {\cal K}}=0 \;.
\end{equation}
Let us 
expand  $s(t, {\bf x})$ in spacial Fourier modes $s_k (t)$, where $k$ is the comoving 3--momentum. If $V$ is quadratic    in $s$, the different momentum modes decouple and we have 
\begin{equation}
\ddot s_k + \left(   {\dot {\cal K} \over {\cal K}} +3H  \right)  \dot s_k +  \left(  {k^2\over a^2}  +  {{\cal V}^{\prime\prime}_s \over {\cal K}} \right) s_k=0 \;.
\end{equation}
Now 
suppose that the renormalizable potential vanishes and 
only one dim--6 operator is present at a time, such that either $ {\cal K} =1$ or ${\cal V}=0$.

{\bf (a)}
Consider first the case of a non--trivial  kinetic function ${\cal K}$ and ${\cal V}=0$. The first order time derivatives can be eliminated by the change of variables
\begin{equation}
s_k = a^{-3/2} {\cal K}^{-1/2}  X_k \;,
\end{equation}  
such that
\begin{equation}
\ddot X_k + \left[    {1\over 4} {    \dot   {\cal K}^2 \over {\cal K}^2  }    - {1\over 2} {    \ddot   {\cal K} \over {\cal K}  }   - {3\over 2} H {    \dot   {\cal K} \over {\cal K}  }
+ {9\over 4} w H^2 + {k^2 \over a^2} 
         \right] X_k=0 \;,
\end{equation}
where the equation of state coefficient is 
\begin{equation}
w= -\left(    1+ {2 \dot H \over 3 H^2}  \right)\;.
\end{equation}
In the limit of a constant inflaton amplitude,  the coefficients are periodic in time such that  the above EOM belongs to the class of {\it Hill's equations}. Depending on the parameters,
the solution $X_k$ can grow exponentially in time signifying particle production.

Let us now specialize to  operator $ {\cal O}_3$,  
\begin{equation}
 {\cal K} = 1+ {2 C_3 \over M_{\rm Pl}^2} \, \phi^2 \;.
 \end{equation}
 The effective field theory expansion makes sense if $C_3 \phi^2/ M_{\rm Pl}^2 \ll 1$.
 Consider the  $locally$ quadratic inflaton potential such that 
  $\phi(t) = \phi_0(t)  \cos m_\phi t$ with $\phi_0 \propto 1/(m_\phi t)$. In this case, the Universe is matter dominated, $w=0$ and 
  \begin{equation}
  H = {m _\phi \phi_0 \over \sqrt{6} M_{\rm Pl}} \;.
  \label{av-H}
  \end{equation}
For $C_3 <1$ and $\phi_0 < M_{\rm Pl}$, the term $(\dot {\cal K}/ {\cal K})^2 $ in the square brackets is insignificant, as is $w H^2$. The terms $H \dot {\cal K}/ {\cal K}$ and 
$\ddot {\cal K}/ {\cal K}$ are similar in magnitude initially, but the former is cubic in $\phi_0$ and thus decreases faster in time.
Keeping just the $\ddot {\cal K}/ {\cal K}$ term, we may approximate
\begin{equation}
\ddot X_k + \left[    { 2 C_3 m_\phi^2  \phi_0^2 \over M_{\rm Pl}^2 } \, \cos 2 m_\phi t + {k^2 \over a^2} 
         \right] X_k=0 \;.
\end{equation}
This has the form of the {\it Mathieu equation},
\begin{equation}
X_k^{\prime\prime} +\left[ A+ 2q \, \cos 4z \right] \, X_k=0 \;,
\end{equation}
where $z= m_\phi t /2$, the prime stands for differentiation with respect to $z$, and 
\begin{equation}
q= {4 C_3 \phi_0^2 \over M_{\rm Pl}^2} ~~,~~ A= {4 k^2 \over a^2 m_\phi^2} \;.
\end{equation}
A large $q$ generally implies fast amplitude growth and efficient particle production via broad parametric resonance \cite{Kofman:1997yn}.
In our case, however,  $q\ll 1$  and the resonance is narrow. As  a result, no efficient particle production is possible, especially in view of the redshifting of the 
produced particle momenta \cite{Kofman:1997yn},\cite{Mukhanov:2005sc}.

 It is important to note that, in the above derivation, we have used the average Hubble rate (\ref{av-H}), as is common. However, the exact $H$ contains a subleading oscillating term 
$\delta H \sim \phi \dot \phi /M_{\rm Pl}^2$ \cite{Ema:2015dka}, whose size is suppressed by $\phi / M_{\rm Pl}$ compared to that of the average $H$. 
The main effect of this term is a correction to $w$, such that $w H^2$ no longer vanishes. This effectively renormalizes  $C_3$ by shifting it by an ${\cal O}(1)$ constant. 
Thus, the conclusion that no broad resonance ($q\gg 1$) is possible still holds, while some amount of dark matter is generated through  this effect perturbatively, along the lines of Section 3
(see Eq.\,\ref{Gamma-ss-1}).

As noted in  \cite{Ema:2015dka}, the effect of the oscillating term $\delta H$ is largely equivalent to that of operator $\phi^2  (\partial_\mu s)^2$.  This can be seen using the EOM for the scale factor and expanding
it in terms of the average and oscillating parts in the action. Thus, on-shell, ${\cal O}_3$ is generated already by classical gravity. One expects that there are also other fundamental sources inducing 
 ${\cal O}_3$  off-shell   such as quantum gravity, string states, etc., whose effect we parametrize in terms of $C_3$.

{\bf (b)}
Let us now consider the effect of ${\cal O}_4 $, so we take ${\cal K}=1$ and 
\begin{equation}
{\cal V}= {C_4 \over M_{\rm Pl}^2 } \; \phi^4 s^2 \;.
\end{equation}
For the quadratic inflaton potential,  the corresponding $X_k$ satisfies
\begin{equation}
\ddot X_k + \left[    { 2 C_4  \phi_0^4 \over M_{\rm Pl}^2 } \, \cos^4 m_\phi t + {k^2 \over a^2} 
         \right] X_k=0 \;.
\end{equation}
Using $\cos^4 x = {1\over 8} \cos 4x + {1\over 2} \cos 2x +{3\over 8}$, one can bring the EOM into the form
of the {\it Whittaker--Hill equation},
\begin{equation}
 X_k^{\prime\prime} + \left[    A + 2p \cos 2z + 2q \cos 4z
         \right] X_k=0 \;,
\end{equation}
where now  $z= m_\phi t $ and 
\begin{equation}
q= { C_4 \phi_0^4 \over 8 m_\phi^2 M_{\rm Pl}^2} ~~,~~ p= { C_4 \phi_0^4 \over 2 m_\phi^2 M_{\rm Pl}^2} ~~,~~
A= { k^2 \over a^2 m_\phi^2} +  {3 C_4 \phi_0^4 \over 4 m_\phi^2 M_{\rm Pl}^2} \;.
\end{equation}
The efficiency of particle production is characterized by $p$ and $q$, whose large values (depending on $A$) generally lead to broad  resonance
\cite{Dufaux:2006ee},\cite{Enqvist:2016mqj}. We see that 
this regime is easily achieved in the presence of ${\cal O}_4 $. In particular, 
$p,q\gg 1$  is consistent with $C_4 \ll 1$ and sub--Planckian $\phi_0$ values, as long as $\phi_0 \gg m_\phi $.

We note that the effect of the oscillating term in the Hubble rate \cite{Ema:2015dka} is insignificant here: the corresponding $q$ is at most order 1, as mentioned earlier. Also, according to the argument of 
Ref.\,\cite{Ema:2015dka}, 
the effective interaction $\phi^4 s^2$ does get  induced in the on-shell action via   the scale factor expansion at higher order, yet its coefficient is suppressed by $m_s^2/M_{\rm Pl}^2$ and can be neglected.

We  conclude that ${\cal O}_4 $ is much more efficient in particle production  than ${\cal O}_3 $.
The same  conclusion applies to the quartic inflaton potential: the analysis proceeds analogously up to the replacement of the inflaton mass with the effective inflaton mass 
$m_\phi^{\rm eff }\sim \sqrt{\lambda_\phi} \phi_0$.

The resonance efficiency is determined by  the ratio of the inflaton--induced DM mass to the inflaton effective mass. 
In the list (\ref{L}), this 
 ratio can be large only for operator ${\cal O}_4 $. Indeed, $(\partial_\mu \phi)^2 s^2$ induces the DM mass of order $m_\phi$ at best. The term 
$(\phi \partial_\mu \phi) (s \partial^\mu s) = (\partial_\mu \phi^2 ) (\partial^\mu s^2)/4 $ can be rewritten as a mass term for $s$ by integrating by parts.
The resulting DM mass scale is determined by $m_\phi$ or $H$, leading to the same conclusion.
The operator $\phi^2 s^4$ does not induce any mass in our approximation and the corresponding EOM is not of Hill's type (see below).\footnote{The induced mass term appears when the variance $\langle s^2 \rangle$ becomes significant.}
Finally, the renormalizable term $\phi^2 s^2$  of the form (\ref{O-ren}) does not lead to broad resonance since the corresponding $q\lesssim 1$.
 It is thus clear that ${\cal O}_4 $ dominates particle production.

Similar conclusions apply to higher dimensional operators $\phi^6 s^2$, etc. As long as $\phi_0$ is not much below the Planck scale, the effective $q$ parameter can be much greater than one, signifying efficient particle production. Therefore, the results are sensitive to the presence of operators of this type.

The amount of   dark matter produced during preheating is difficult to estimate analytically. The reason is that the parameters of the Whittaker--Hill equation evolve in time making the resonance stochastic,
which is further complicated by the non--trivial 3-D stability band structure \cite{Enqvist:2016mqj}. 
Depending on the size of $C_4$, tangible backreaction and rescattering effects  \cite{Khlebnikov:1996mc},\cite{Prokopec:1996rr} can also take place. Thus,
 to make reliable predictions, we have to resort to lattice simulations.

\subsection{On resonant production via $\phi^2 s^4$} 

Unlike for other operators considered in this work, 
resonant particle production
via  $ {\cal O}_5 = \phi^2 s^4$ is not described by the Hill's equation. 
In this case,
the system is   non--linear already to leading order and thus difficult to handle analytically. In this subsection, we discuss some of its properties  relevant to the subject of our paper.

The EOM for the DM field  $s$ in the presence of  $ {\cal O}_5 $ reads
\begin{equation}
\ddot s - {1\over a^2} \,\partial_i \partial_i s  +3H\, \dot s + {4 C_5 \phi^2 (t) \over M_{\rm Pl}^2} \, s^3 =0 \;.
\end{equation}
Clearly, the different momentum modes do not decouple in this case. As a representative example, let us focus on the zero mode of $s$ which normally plays a major role in particle production.
Omitting the gradient term and introducing a rescaled field $X$ ($s=a^{-3/2} X$), we get 
\begin{equation}
\ddot X  + {9\over 4 } w H^2 \, X + {4 C_5 \phi^2 (t) \over M_{\rm Pl}^2 \,a^3} \, X^3 =0 \;,
\end{equation}
where $w$ is the coefficient of the equation of state of the system.
Let us now specialize to the quadratic inflaton potential such that $w=0$,
\begin{equation}
{a\over a_0} = \left( {m_\phi t \over m_\phi t_0}\right)^{2/3} \;,
\end{equation}
and $\phi (t) \simeq 1.85  \,\varphi_0 {\cos \, m_\phi t \over m_\phi t}$. Here the initial condition is chosen such that $\phi (t_0) = \varphi_0$ with $m_\phi t_0=1$.
The above EOM should be supplemented by boundary conditions. The magnitude of a light  field is given by quantum fluctuations, such that we may take $s\sim H$ and 
$\dot s \sim H^2$ initially, where $H = m_\phi \varphi_0 /(\sqrt{6} M_{\rm Pl})$.   Introducing 
\begin{equation}
z= m_\phi t ~~,~~ Y = X/m_\phi \;,
\end{equation}
we get 
\begin{equation}
Y^{\prime \prime} + \kappa \, {\cos^2 z \over z^4} \, Y^3 =0 \;,
\label{Y-eq}
\end{equation}
where the prime denotes differentiation with respect to $z$. The representative boundary conditions can be chosen as $Y(1)=1, Y^\prime(1)=1$. The coupling $\kappa$ is given by 
\begin{equation}
\kappa \simeq {14 C_5 \varphi_0^2 \over M_{\rm Pl}^2\, a_0^3} \;,
\end{equation}
where typically $\varphi_0 \sim M_{\rm Pl}$, $a_0 \sim 1$ and the effective field theory approach is expected to be valid for $\kappa \lesssim {\cal O}(1)$.

The oscillating term in Eq.\,\ref{Y-eq} falls faster with time then the analogous coefficient in the Mathieu equation, hence the duration of the resonance is shorter in the present case.
The equation exhibits a simple asymptotic solution for $z\gg 1$  (or $\kappa \ll 1$),
\begin{equation}
Y \propto z \;.
\label{asym}
\end{equation}
On the other hand, for $\kappa \gg 1$,  $Y(z)$ varies much faster than $\cos^2z/z^4$ does such that the latter can be treated adiabatically. In this case, 
Eq.\,\ref{Y-eq} takes the form $Y^{\prime \prime} + c \, Y^3 =0 $ with $c= \kappa \, \cos^2z/z^4$, whose solution is a Jacobi cosine. 
Given the amplitude of oscillations $Y_0$, locally we have 
 \begin{equation}
Y(z) \simeq Y_0 \, {\rm cn} \left(\sqrt{c} \,Y_0 z, {1\over \sqrt{2}}\right) 
\end{equation}
 in the convention of \cite{Greene:1997fu}. The oscillation frequency $\sqrt{c} Y_0$ changes non--adiabatically around the inflaton zero crossings, implying particle production.

    \begin{figure}[h!] 
\centering{
\includegraphics[scale=0.36]{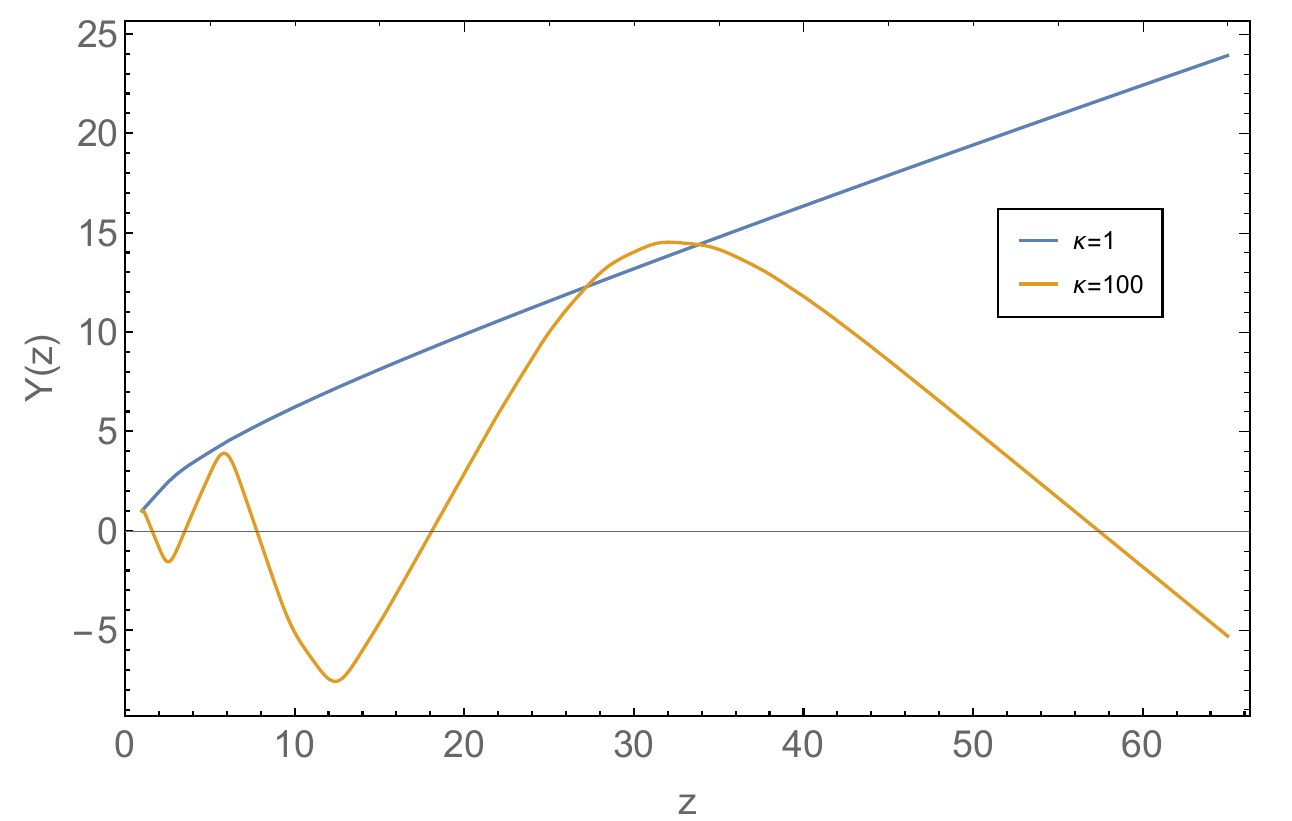}
\includegraphics[scale=0.36]{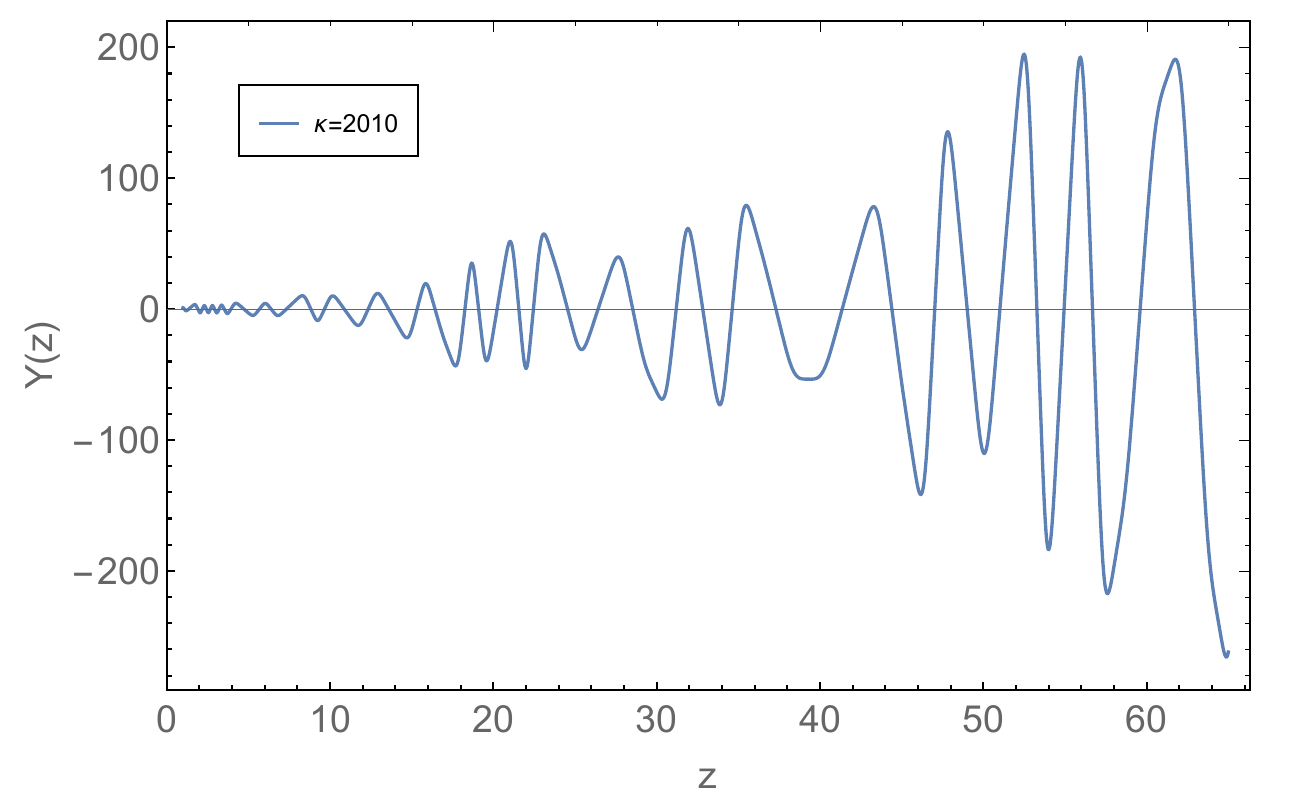}
}
\caption{ \label{Y(z)}
Solutions to Eq.\,\ref{Y-eq} for $\kappa =1, 100, 2010$.}
\end{figure}

Numerical solutions to Eq.\,\ref{Y-eq} are presented in Fig.\,\ref{Y(z)}. For $\kappa =1$, the solution quickly takes on the asymptotic form (\ref{asym}), which corresponds to constant $s$
in our approximation ($m_s \sim 0$). 
Even for $\kappa =100$, there is no significant 
amplitude growth, while for $\kappa =2010$ the solution exhibits truly resonant behaviour.\footnote{We find that the solution in this regime is  quite sensitive to the numerical method used 
 for solving the differential equation. }
 We thus find that  $\kappa \gtrsim 10^3 $ is necessary for efficient DM production.

The discussion above concerns the zero mode. Normally it serves as an indicator whether or not resonant particle production takes place. To reaffirm it, we have 
performed lattice simulations of the full inflaton--DM system in the regime  $C_5 \phi^2 /M_{\rm Pl}^2 \lesssim 1$ and indeed found no tangible particle production.

As before, we have used the average Hubble rate and neglected its oscillatory component \cite{Ema:2015dka}. The latter induces a non--zero oscillating $wH^2 $ term, which however does 
not lead to efficient particle production. Although some amount of dark matter gets produced perturbatively as a result, this does not obscure our analysis  of the $C_5$-induced effects.

We  conclude that no efficient particle production is induced by operator ${\cal O}_5$ as long as   $C_5 \phi^2 /M_{\rm Pl}^2 \lesssim  {\cal O}(1)$.

\section{Lattice simulations: reproducing the correct DM abundance}
\label{Simulation}

In this Section, we focus on the leading operator $ {C_4 \over M_{\rm Pl}^2}\,\phi^4 s^2$ and compute the resulting DM relic abundance in the resonant regime, $C_4 \phi_0^4 /M_{\rm Pl}^2 \gg m_\phi^2$,
assuming $m_\phi \gg m_s$. As explained above, the analytical approach to resonant particle production has  significant limitations,
given the complexity of the Whittaker--Hill equation as well as backreaction and rescattering effects. Therefore, we resort to lattice simulations using the numerical tool CosmoLattice \cite{Figueroa:2021yhd,Figueroa:2020rrl}.

A realistic framework must also account for the Standard Model  particle production. The simplest way to incorporate reheating is to include a small Higgs--inflaton coupling
following \cite{Lebedev:2021xey,Lebedev:2021tas},
\begin{equation}
V_{\phi h} =  \sigma_{\phi h} \phi H^\dagger H \;,
\end{equation}
 which would lead to late--time decay of the inflaton into the Higgs pairs (for $m_\phi > 2 m_h$). As long as $ \sigma_{\phi h} $ is sufficiently small in Planck units, resonant dark matter production 
 is not affected by this coupling.
 The dark matter abundance is  expressed in terms of 
    \begin{equation}
    Y = {n \over s_{\rm SM} } ~~,~~ s_{\rm SM} ={2\pi^2 \over 45} \, g_{*s} \, T^3 \;,
    \label{Y}
    \end{equation} 
    where  $n$ is the DM number {} density, $s_{\rm SM}$ is the Standard Model entropy {} density at temperature $T$ and 
    $g_{*s}$ is the effective number of SM degrees {} of freedom contributing {} to the entropy. 
    We are interested in very weakly interacting dark matter such that it never reaches thermal equilibrium with the environment.
    After preheating ends, the total number of the DM quanta remains constant. Since the SM entropy is also conserved, $Y$ 
    can be computed at the reheating stage.
     The observed value is   \cite{Planck:2015fie}   
     \begin{equation}
    Y_\infty =  4.4 \times 10^{-10} \; \left( {{\rm GeV}\over m_s} \right) \;,
    \end{equation} 
which sets a constraint on the model parameters.

 Reheating occurs when 
  \begin{equation}\label{reh_cond}
  H_R \simeq \Gamma_{\phi\rightarrow hh} \;,\quad
  \Gamma_{\phi\rightarrow hh}=\frac{\sigma_{\phi h}^2}{8\pi m_\phi},
\end{equation}
where \(H_R\) is the Hubble rate at  reheating and   \(\Gamma_{\phi\rightarrow hh}\)
takes into account 4 Higgs d.o.f.   at high energies. The reheating temperature is given by
\begin{equation}
  H_R=\sqrt{\frac{\pi^2g_*}{90}} \, {T_R^2 \over M_{\rm Pl}} \;,
  \label{HR}
\end{equation}
where $g_* $  is the effective number of the Standard Model degrees {} of freedom contributing {} to the energy density.
Combining $T_R$ with the dark matter density $n$ computed on the lattice, one determines $Y$ according to (\ref{Y}).
  
The resulting abundance  is sensitive to  the energy balance between the inflaton and dark matter, which affects the expansion history.    Depending on the coupling and the inflaton potential, 
DM can contribute a significant fraction up to 50\% to the total energy density at the end of preheating. We therefore parametrize
\begin{equation}
  \rho_e (s)=\delta \, \rho_e (\phi) \;,
  \end{equation}
  where $\rho_e (s)$, $\rho_e (\phi)$ are the DM and inflaton energy densities, respectively, evaluated  at the end of the simulation. 
  At weak coupling, in the quadratic inflaton potential we have $\delta \sim 0$, while, at strong coupling, $\delta$ can reach a value close to 1.
   The Universe evolution proceeds in       stages: first, it can be dominated by radiation; later, when the energy per quantum  becomes comparable to the inflaton mass, it evolves
   as non--relativistic matter; finally, the Universe reheats and becomes radiation--like. 
Denoting the corresponding scale factors as $a_e$ (end of the simulation), $a_*$ (transition), $a_R$ (reheating), we have 
    \begin{equation}
a_e  \stackrel{\rm { rel}}  \longrightarrow         a_*     \stackrel{\rm nrel} \longrightarrow a_R \;,
\end{equation} 
   such that  the Hubble rate  evolves  as $H\sim a^{-3(w+1)/2}$ with $w=1/3$ and $w=0$ during the two periods, respectively.
  After the  transition point $a_*$, $\rho (s)$ becomes negligible and at $a_R$  the inflaton energy density $\rho (\phi)$ converts into SM radiation.  Thus,
   \begin{equation}
  H_R=\frac{H_e}{\sqrt{1+\delta}} \, \frac{a_e^2}{a_*^2} \, \frac{a_*^{3/2}}{a_R^{3/2}} \; ,
\end{equation}
where $H_e$ is the Hubble rate at the end of the simulation. 
We note  that the first stage of the radiation--like expansion may collapse to a point, i.e.
   $a_e = a_*$. This is the case  for the quadratic inflaton potential at weak inflaton--DM coupling.  

Solving for $a_R$, we find $\sigma_{\phi h}$ required by the correct DM abundance in terms of the simulation output:
 \begin{equation}
\sigma_{\phi h} \simeq  1.6 \times 10^{-8} \; \sqrt{m_\phi \,M_{\rm Pl}^3}  \; {H_e^2 \over (1+\delta) \;n_e} \, {a_e\over a_*} ~~\left(   {{\rm GeV}  \over m_s} \right) 
\label{sigma-DM-eq}
\end{equation}
for  $g_* \simeq 107$ and $M_{\rm Pl}$ being the reduced Planck mass.
The values of   $H_e, n_e, \delta$  at the end of preheating   are computed  by CosmoLattice, while $a_e/a_*$ can be determined by tracking the equation of state of the system.

This formula can be simplified further if we define $a_*$ according to 
\begin{equation}
 \langle E_e (\phi) \rangle \; {a_e \over  a_*} \simeq m_\phi \;,
\end{equation} 
   where  the average energy of the inflaton quantum at the end of the simulation is $ \langle E_e (\phi) \rangle = \rho_e (\phi) / n_e (\phi)$. 
   This definition of $a_*$ is more practical in that it does not require tracking the equation of state of the system over a long period, which is computationally challenging.
   We then get
  \begin{equation}
\sigma_{\phi h} \simeq 5 \times 10^{-9} \;  {m_\phi^{3/2} \over M_{\rm Pl}^{1/2}}   ~  {n_e(\phi) \over n_e(s)} ~~\left(   {{\rm GeV}  \over m_s} \right) \;, 
\label{sigma-DM-eq-1}
\end{equation}
which only requires the particle densities as an output of the simulations. Here $n_e(\phi)$ includes the inflaton quanta with zero momentum and  
typically $n_e (\phi) \gg n_e (s)$ unless the coupling is relatively strong.

The number densities are computed via the $k$--mode occupation numbers $n_k$. For the dark matter field, we have 
\begin{equation}
n_k = {\omega_k \over 2} \, \left(  {   \vert \dot X_k \vert^2   \over \omega_k^2} + \vert  X_k \vert^2   \right) -{1\over 2} \;,
\end{equation}
  where $\omega_k^2 (t) =  { 2 C_4  \phi_0^4 \over M_{\rm Pl}^2 } \, \cos^4 m_\phi t + {k^2 \over a^2} $ in the quadratic inflaton potential.
  Here $X_k$ is a solution to the EOM with the boundary condition given by  quantum fluctuations. The resulting number density is then given by
  \begin{equation}
n(s) =   {1\over (2\pi a)^3 } \int  d^3 k \;  {n_k } \,.
\end{equation}
    \begin{figure}[h!] 
\centering{
\includegraphics[scale=0.29]{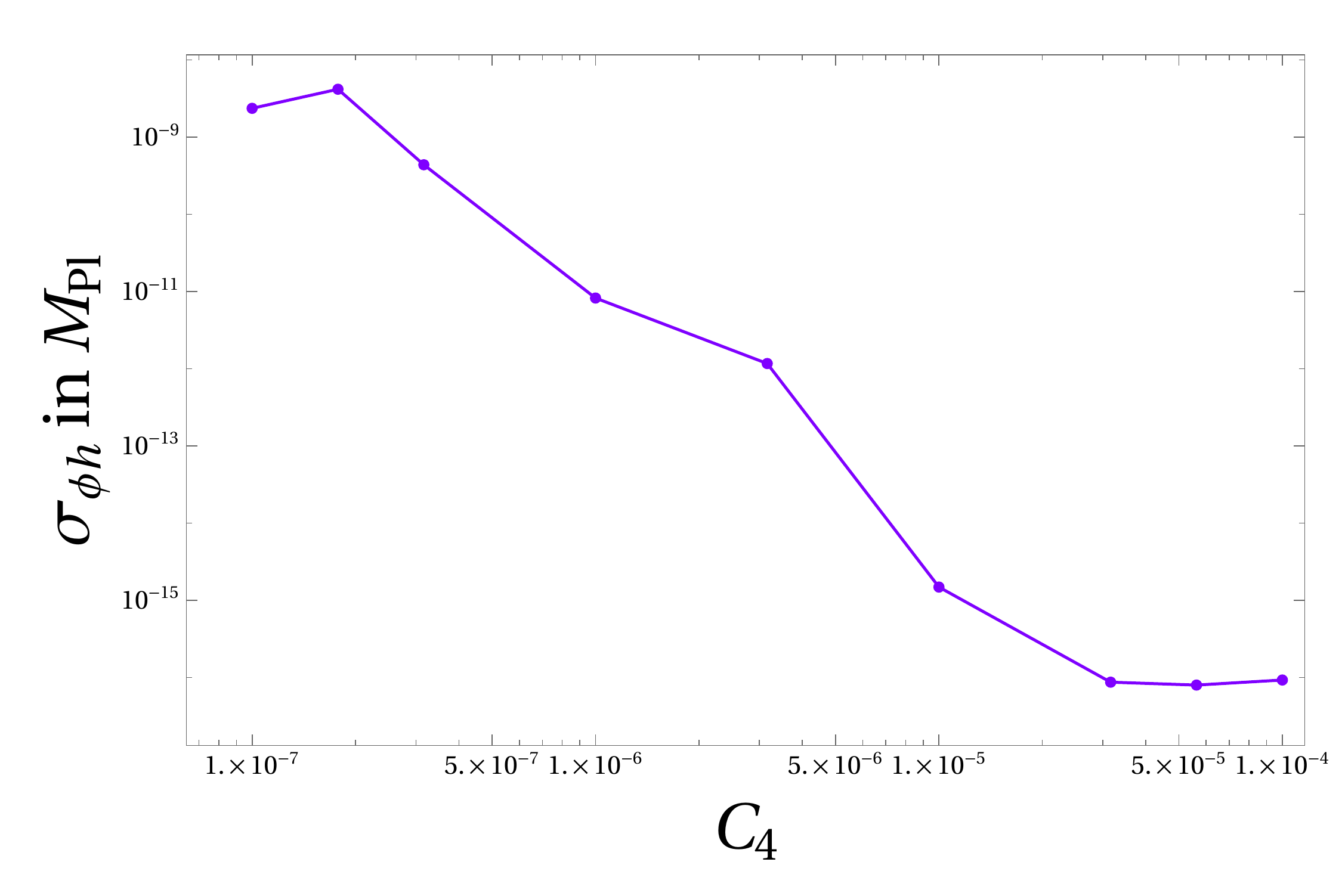}
}
\caption{ \label{C4}
  $C_4$ vs $\sigma_{\phi h }$ required for the correct DM abundance in the quadratic inflaton potential ($m_\phi = 10^{13}$ GeV, $m_s=1$ GeV, $\varphi_0 \simeq M_{\rm Pl}$).
The simulations are performed with CosmoLattice \cite{Figueroa:2021yhd}.
}
\end{figure}
  On the lattice, the momentum spectrum is discrete which allows one to treat the zero mode separately.
  Analogous formulae apply to the inflaton field and the quartic potential. 
   It is important to remember  that the EOM for the different momentum modes decouple at 
  weak couplings only. The lattice approach allows us to incorporate the couplings among the $k$--modes of the inflaton and DM, thereby
  accounting for backreaction and rescattering. The latter can have a crucial impact on the dynamics of the system (see \cite{Lebedev:2021zdh,Kost:2021rbi} for  recent examples).

 Our numerical results are presented in Fig.\,\ref{C4}. The correct relic density is produced in a wide range of $C_4$ between $10^{-7}$ and $10^{-4}$. The lower bound comes from requiring
 semiclassical behaviour, that is, the occupation numbers must be sufficiently large. The upper bound has technical nature: the simulation becomes unstable. We observe that the curve tends to flatten out at larger couplings. This is expected from quasi--equilibrium: as $n_e(\phi)$ approaches $n_e (s)$,  the $\sigma_{\phi h}$ coupling becomes constant within our approximation \cite{Lebedev:2021tas}.  Although such flattening is clearly visible, we find that 
 quasi--equilibrium has not yet been reached at $C_4 \sim 10^{-4}$. We estimate the required $C_4$ to be of order $10^{-3}$, yet the simulations in this range become less reliable.

 Given these results, we can now determine under what circumstances  the renormalizable coupling ${1\over 4}\lambda_{\phi s} \phi^2 s^2 $ becomes unimportant. According to Fig.\,4 of \cite{Lebedev:2021tas},
 $\lambda_{\phi s} < 10^{-8}$ does not make any significant contribution to the dark matter abundance in the parameter range of interest. 
 This can be understood intuitively since $\lambda_{\phi s}/4 < C_4 \varphi^2_0 / M_{\rm Pl}^2$ in this case (see also \cite{Dufaux:2006ee}).
 On the other hand, for 
  $\lambda_{\phi s} > 10^{-7}-10^{-6}$ or $C_4 > 10^{-3}$, the inflaton--DM system reaches quasi--equilibrium and the DM abundance becomes  independent  of these couplings,
  with the required $\sigma_{\phi h} \sim 10^{-17} M_{\rm Pl}$.

  Subsequently, when the inflaton coherence is lost, the operator ${\cal O}_4$ relinquishes  its privileged role in DM production. 
  In quasi--equilibrium, the scattering processes $\phi \phi \rightarrow ss$ can become comparably significant, depending on the corresponding $C_i$. Since all of our operators are Planck--suppressed,
  such processes are slower than the Hubble rate and thus do not lead to inflaton thermalization \cite{Lebedev:2021ixj} nor significant DM production.\footnote{
  For the parameters of the plot, the reheating temperature ranges from $10^{11}$ GeV at low $C_4$ to $10^4$ GeV at larger $C_4$. The corresponding dark matter production via graviton exchange is negligible 
\cite{Mambrini:2021zpp}, while for consistency all the operators suppressed by $1/M_{\rm Pl}^2$ would have to be included in this calculation.} 
  At weak couplings, the inflaton field may remain semi--classical during reheating, in which case the effects of the SM thermal bath can be included along the lines of 
  \cite{Ai:2021gtg},\cite{Wang:2022mvv}.

  Throughout this work we assume that other sources of dark matter are subdominant. 
  In particular,  we neglect  the DM coupling to the Higgs 
   such that no tangible freeze--in contribution from the Higgs thermal bath appears. This approximation is justified if this coupling is below $ 10^{-11}$
   \cite{Lebedev:2019ton}.\footnote{
   In some cases, e.g. when an inverse phase transition in the dark sector is possible, even smaller Higgs portal couplings can produce the right amount of dark matter \cite{Ramazanov:2021eya}.} 
 Furthermore, 
 as explained in the Introduction,
  there is also a truly gravitational source of DM: the Universe expansion. However, in the presence of ${\cal O}_4$, the effective mass of $s$ during inflation can be larger than the Hubble rate 
due to super--Planckian inflaton values, which suppresses $\langle s^2 \rangle$ and makes this production mechanism inefficient.

  We have focused on dim--6 operators, while in general one also expects $Z_2$ breaking dim--5   terms such as  $\phi^3 s^2$. Their effects would be very sensitive to the size of $Z_2$  violation. 
If the DM--inflaton interactions exhibit the same amount of parity violation as the Higgs--inflaton interaction does (quantified by $\sigma_{\phi h}/M_{\rm Pl}$), we expect the right-hand side of Fig.\,2 to be immune
to such couplings since $C_4 $ would be very much larger than the corresponding Wilson coefficient of the dim--5 operator. In general, however, dim--5 operators can dominate particle production.

  In our analysis, we have relied on the effective field theory expansion in the Einstein frame, which is expected to be meaningful during preheating. 
  Gravitational dark matter production can also be encoded in the DM non--minimal coupling to gravity \cite{Fairbairn:2018bsw}, which corresponds to a specific choice of higher dimensional operators in our approach.
  A related option is provided by 
      gravity--induced inflaton decay in Starobinsky--like models \cite{Li:2021fao}.

\section{Conclusion}

In this work, we have studied perturbative and non--perturbative dark matter production during the inflaton oscillation epoch. 
We focus on the regime where 
  the renormalizable interactions between the inflaton and dark matter are negligible.
To determine the leading contributions, we 
     resort to the effective field theory expansion in the inverse Planck mass. Such higher dimensional operators are expected to be generated by perturbative or non--perturbative gravitational effects.
   In the absence of quantum gravity theory, their coefficients   are unknown and therefore treated as arbitrary input parameters. 
   
   We have focussed on Planck--suppressed dim--6 operators and  studied their relative importance in the perturbative and resonant regimes. 
We find that  operators of the form $\phi^n s^2$ ($n \geq 4$) by far dominate particle production. They can generate the correct (non--thermal) dark matter abundance even for small values of the corresponding Wilson coefficients. Therefore, the phenomenological frameworks describing dark matter production are sensitive to the presence of such operators,  
which reinforces 
   the importance of gravitational effects in this context.
   \\ \ \\{\bf Acknowledgements.} The authors wish to thank the Finnish Computing Competence Infrastructure (FCCI) for supporting this project with computational and data storage resources.
   \\ \ \\ \ \\

\end{document}